\documentclass[preprint,showpacs,amsmath,amssymb,aps,prd,nofootinbib]{revtex4}
\usepackage{graphicx,graphics}
\usepackage{amssymb}
\usepackage{amsmath}
\usepackage{dcolumn}
\usepackage{bm}
\usepackage{color}

\newcommand{\bi}{\bibitem}
\def\be{\begin{equation}}
\def\ee{\end{equation}}
\newcommand{\bea}{\begin{eqnarray}}
\newcommand{\eea}{\end{eqnarray}}
\newcommand{\nn}{\nonumber}

\def\hbar#1{\backslash\hspace{-2mm}#1}

\def\nn{\nonumber}

\def\2tvec#1#2{
\left(
\begin{array}{c}
#1  \\
#2  \\
\end{array}
\right)}

\def\mat2#1#2#3#4{
\left(
\begin{array}{cc}
#1 & #2 \\
#3 & #4 \\
\end{array}
\right) }

\def\Mat3#1#2#3#4#5#6#7#8#9{
\left(
\begin{array}{ccc}
#1 & #2 & #3 \\
#4 & #5 & #6 \\
#7 & #8 & #9 \\
\end{array}
\right) }

\def\3tvec#1#2#3{
\left(
\begin{array}{c}
#1  \\
#2  \\
#3  \\
\end{array}
\right)}

\def\hbar#1{\backslash\hspace{-2mm}#1}

\def\nn{\nonumber}

\newcommand{\bt}{\begin{itemize}}
\newcommand{\et}{\end{itemize}}

\numberwithin{equation}{section}

\begin{document}

\begin{flushright}
KIAS-P13014
IPPP/13/14
DCPT/13/28
\end{flushright}

\begin{center}
{\large \bf Multicomponent dark matter particles in a two-loop neutrino model}
\end{center}

  \author{Yuji Kajiyama \footnote{E-mail: kajiyama-yuuji@akita-pref.ed.jp}}
 \author{Hiroshi Okada \footnote{E-mail: hokada@kias.re.kr}}
 \author{Takashi Toma \footnote{E-mail: takashi.toma@durham.ac.uk}}
\

 \affiliation{Akita Highschool, Tegata-Nakadai 1, Akita, 010-0851, Japan}
 \affiliation{ School of Physics, KIAS, Seoul 130-722, Korea}
 \affiliation{ Institute for Particle Physics Phenomenology University of Durham, Durham DH1 3LE, UK}

\date{\today}

\begin{abstract}
We construct a loop induced seesaw model in a TeV scale theory with
 gauged $U(1)_{B-L}$ symmetry. Light neutrino masses are generated
 at two-loop level and right-handed neutrinos also obtain their masses
 by one-loop effect. Multi-component Dark Matters (DMs) are included in our 
model due to the remnant discrete symmetry after the $B-L$ symmetry
 breaking and the $\mathbb{Z}_2$ parity which is originally imposed to the model. We investigate the multi-component DM properties, 
in which we have two fermionic DMs with different  mass
 scales, 
${\cal O}$(10) GeV and ${\cal O}$ (100-1000) GeV. 
 The former mass corresponds to the lightest right-handed neutrino mass induced by the loop effect,
although the latter one to the SM gauge singlet fermion. 
We show each of the DM annihilation processes and compare to the observation of relic abundance,
together with the constraints of Lepton Flavor Violation (LFV) and active neutrino masses. 
Moreover we show that our model has some parameter region allowed by the direct detection result reported by XENON100,
and it is possible to search the region by the future XENON experiment. 
\end{abstract}

\maketitle

\section{Introduction}
It has been verified that neutrinos have tiny masses by neutrino
oscillation experiments~\cite{Abe:2011fz,An:2012eh,Ahn:2012nd,Apollonio:2002gd,Boehm:2001ik,Abe:2011sj,Adamson:2011qu}.
Unfortunately the finite neutrino masses
are not explained in the framework of the Standard Model (SM). A lot of
models have been proposed to extend this point~\cite{seesaw,Khalil:2006yi,Mohapatra:1986bd,GonzalezGarcia:1988rw}. 
On the other hand the existence of non-baryonic Dark Matter (DM),
which dominates about 23\% of the Universe from the CMB observation by WMAP \cite{wmap},
is also shown by the cosmological
observations in our Universe~\cite{Begeman:1991iy,Massey:2007wb}
{\footnote{Very recently, new result of the DM relic density was 
given by Planck measurement as $\Omega_C h^2=0.1196 \pm 0.0031$\cite{Ade:2013lta}.}}. 
Moreover direct detection experiments of DM are performed around the
world such as XENON100~\cite{xenon100}, CRESSTII~\cite{cresst},
CoGeNT~\cite{cogent}, DAMA~\cite{dama} and TEXONO~\cite{Li:2013fla}. 
Especially XENON100 experiment gives the most severe limit for 
elastic scattering cross section between DM and
nucleon~\cite{xenon100}. 
This implies that DM in the Universe interacts
very weakly with quarks. It would be that DM has no interaction with quarks.
DM is required in the Universe, however a candidate particle
of DM is also not included in the
SM. Although the property of neutrino in the SM is similar with that of
DM, neutrinos are too light to be DM candidate. 
Thus in order to improve this problem, it is necessary to add 
new particles as DM candidate in the SM. 
Therefore, these current experiments about neutrinos and DM suggest serious
verifications that the SM should be modified in order to accommodate
the existence of DM as well as non-vanishing neutrino masses. 

Radiative seesaw models are known as attractive frameworks for new
physics at TeV scale that can provide an elegant solution to explain
these two matters of grave concern simultaneously~\cite{Ma:2006km,
Aoki:2013gzs, Krauss:2002px, Aoki:2008av}. 
This kind of model correlates the finite neutrino masses with the
existence of DM since neutrino masses are generated by radiative effect
and DM runs inside the loop. 
In particular, the radiative seesaw model proposed by Ernest Ma
\cite{Ma:2006km} is one of the simplest models. Subsequently there are a
lot of recent works in terms of the model~\cite{Schmidt:2012yg, Bouchand:2012dx,
Ma:2012ez}
and the extended models \cite{Suematsu:2010nd, Aoki:2011he, Ahn:2012cg,
Farzan:2012sa, Bonnet:2012kz, Kumericki:2012bf, 
Kumericki:2012bh, Ma:2012if, Gil:2012ya, Okada:2012np, Hehn:2012kz,
Dev:2012sg, Kajiyama:2012xg, Okada:2012sp}. 
The other models of radiative neutrino mass are studied in
Refs.~\cite{Aoki:2010ib, Kanemura:2011vm, Lindner:2011it, 
Kanemura:2011mw, two-triplet,Gu:2007ug,Gu:2008zf,Gustafsson:2012vj,Law:2013dya, Gu:2013nya}. 

In this paper, we propose a new model of two-loop induced neutrino
masses with local $B-L$ symmetry. 
Due to the two remnant Abelian symmetries ($\mathbb{Z}_2$ and
$\mathbb{Z}_6$) even after the $B-L$ and electroweak symmetry breaking,
our model has multi-component DMs. 
Two or three particles of them can be DMs simultaneously depending on
the mass hierarchy. 
Moreover since one of DMs obtains the mass at one loop, we expect it to
be rather light with mass of ${\cal O}(10)$ GeV. 
We check whether they can satisfy the correct relic density of
DM observed by WMAP, and also the upper bound of elastic cross section
with nucleon by XENON100.

This paper is organized as follows. In Section 2, we show our model and
discuss the Higgs sector including the Higgs potential,  stationary
condition, S-T parameters and neutrino mass in lepton sector.
In Section 3, we analyze DM phenomenologies. We summarize and conclude
in Section 4.

\section{The Two-loop Radiative Seesaw Model}
\subsection{Model setup}

\begin{table}[thbp]
\centering {\fontsize{10}{12}
\begin{tabular}{||c|c|c|c|c|c|c||c|c||}
\hline\hline ~~Particle~~ & ~~$Q$~~ & ~~$u^c$ & $ d^c $~~ & ~~$L$~~ & ~~$e^c$~~ & ~~ $N^c$~~
  & $S$  & $\bar{S}$ \\\hline
$(SU(2)_L,U(1)_Y)$ & $(\bm{2},1/6)$ & $(\bm{1},-2/3)$ & $(\bm{1},1/3)$ & $(\bm{2},-1/2)$ & $(\bm{1},1)$  & $(\bm{1},0)$  & $(\bm{1},0)$ & $(\bm{1},0)$\\\hline
$Y_{B-L}$ & $1/3$ & $-1/3$ & $-1/3$ & $-1$ & $1$ & $1$  & $-1/2$ & $1/2$ \\\hline
$\mathbb{Z}_2$ & $+$ & $+$ & $+$ & $+$ & $+$  & $-$  & $-$ & $+$\\
\hline
$\mathbb{Z}_6$ & $2$ & $-2$ & $-2$ & $0$ & $0$  & $0$ & $-3$ & $3$\\
\hline\hline
\end{tabular}%
} \caption{The particle contents and the charges for fermions.
 Notice that the $\mathbb{Z}_6$ is the remnant symmetry obtained after
 $B-L$ symmetry breaking as we will discuss later.}
\label{tab:b-l1}
\end{table}

\begin{table}[thbp]
\centering {\fontsize{10}{12}
\begin{tabular}{||c|c|c|c|c||}
\hline\hline ~~Particle~~ & ~~$\Sigma$~~ & ~~$\Phi~~ $& ~~$\eta~~ $ & $\chi$ \\\hline
$(SU(2)_L,U(1)_Y)$ & $(\bm{1},0)$ &  $(\bm{2},1/2)$ & $(\bm{2},1/2)$ & $(\bm{1},0)$ 
\\\hline
$Y_{B-L}$ &  $1$ & $0$ & $0$ & $-1/2$  \\\hline
$\mathbb{Z}_2$ & $+$ & $+$ & $-$ & $+$  \\\hline
$\mathbb{Z}_6$ &  $0$ & $0$ & $0$ & $-3$ \\
\hline\hline
\end{tabular}%
} \caption{The particle contents and the charges for bosons. Notice
 that the $\mathbb{Z}_6$ is the remnant symmetry obtained after
 $B-L$ symmetry breaking as we will discuss later.}
\label{tab:b-l2}
\end{table}

We propose a two-loop radiative  seesaw model with $U(1)_{B-L}$ 
which is an extended model of the seesaw model of Ma~\cite{Ma:2006km}. 
The particle contents are shown in Tabs.~\ref{tab:b-l1} and 
\ref{tab:b-l2}. We add three
right-handed neutrinos $N^c$, three SM gauge singlet fermions $S$ and $\bar S$, 
a $SU(2)_L$ doublet scalar $\eta$ and $B-L$ charged scalars $\chi$ and
$\Sigma$ to the SM content, where
 $\eta$ and $\chi$ are assumed not to have vacuum expectation value (VEV). 
The $B-L$ charged scalar $\Sigma$ is the source of the spontaneous $B-L$ 
breaking by its VEV of 
$\langle \Sigma \rangle=v'/\sqrt{2}\sim O(10)$ TeV.
The $\mathbb{Z}_2$ parity is also imposed so as to stabilize DM
candidates.
The right-handed neutrinos $N^c$ do not have masses at tree level.
As a result, the neutrino mass is obtained not through the one-loop
level (just like Ma-model ~\cite{Ma:2006km})
but through the two-loop level. 

The renormalizable Lagrangian for Yukawa sector and Higgs potential
are given by
\begin{eqnarray}
-\mathcal{L}_{\mathrm{Yukawa}}
&=&
y_{\ell}\Phi^\dag e^c L + y_{\nu}\eta^\dag L N^c + y_N N^c\chi S +
y_S\Sigma SS  + y_{\bar{S}}\Sigma^\dag\bar{S}\bar{S}
+\mathrm{h.c.},\\
-\mathcal{L}_{\mathrm{Higgs}}
&=& 
 m_1^{2} \Phi^\dagger \Phi + m_2^{2} \eta^\dagger \eta  + m_3^{2}
 \Sigma^\dagger \Sigma  + m_4^{2} \chi^\dagger\chi 
 + m_5 [\chi^2 \Sigma + {\rm h.c.}]  \nonumber \\ 
&&
+\lambda_1 (\Phi^\dagger \Phi)^{2} + \lambda_2 
(\eta^\dagger \eta)^{2} + \lambda_3 (\Phi^\dagger \Phi)(\eta^\dagger \eta)
+ \lambda_4 (\Phi^\dagger \eta)(\eta^\dagger \Phi)
+\lambda_5 [(\Phi^\dagger \eta)^{2} + \mathrm{h.c.}]\nn\\
&&+\lambda_6 (\Sigma^\dagger \Sigma)^{2} + \lambda_7  (\Sigma^\dagger \Sigma)(\Phi^\dagger \Phi)
+ \lambda_8  (\Sigma^\dagger \Sigma) (\eta^\dagger \eta)+\lambda_9 (\chi^\dagger \chi)^{2}\nn\\
&& + \lambda_{10} (\chi^\dagger \chi)(\Phi^\dagger \Phi)
+ \lambda_{11} (\chi^\dagger \chi) (\eta^\dagger \eta) + \lambda_{12} |\chi \Sigma|^2 ,
\label{HP}
\end{eqnarray}
where $\lambda_5$ has been chosen real without any loss of
generality. The couplings $\lambda_1$, $\lambda_2$, $\lambda_6$ and
$\lambda_9$ have to be positive
to stabilize the Higgs potential.  
Inserting the tadpole conditions; $m^2_1=-\lambda_1v^2-\lambda_7v'^2/2$ and
 $m^2_3=-\lambda_6v'^2 - \lambda_7v^2/2$,
the resulting mass matrix of the neutral component of $\Phi$ and 
$\Sigma$ defined as 
\begin{equation}
\Phi^0=\frac{v+\phi^0(x)}{\sqrt{2}},\qquad
\Sigma=\frac{v'+\sigma(x)}{\sqrt{2}},\quad
\end{equation}
 is given by
\begin{equation}
m^{2} (\phi^{0},\sigma) = \left(%
\begin{array}{cc}
  2\lambda_1v^2 & \lambda_7vv' \\
  \lambda_7vv' & 2\lambda_6v'^2 \\
\end{array}%
\right) \!=\! \left(\begin{array}{cc} \cos\alpha & \sin\alpha \\ -\sin\alpha & \cos\alpha \end{array}\right)
\left(\begin{array}{cc} m^2_{h} & 0 \\ 0 & m^2_{H}  \end{array}\right)
\left(\begin{array}{cc} \cos\alpha & -\sin\alpha \\ \sin\alpha &
      \cos\alpha \end{array}\right), 
\end{equation}
where $h$ implies SM-like Higgs and $H$ is an additional Higgs mass
eigenstate. The mixing angle $\alpha$ is given by 
\be
\tan 2\alpha=\frac{\lambda_7 v v'}{\lambda_1 v^2-\lambda_6 v'^2}.
\ee
The Higgs bosons $\phi^0$ and $\sigma$ are rewritten in terms of the mass eigenstates $h$ and $H$ as
\begin{eqnarray}
\phi^0 &=& h\cos\alpha + H\sin\alpha, \nn\\
\sigma &=&- h\sin\alpha + H\cos\alpha.
\label{eq:mass_weak}
\end{eqnarray}
The other scalar masses are found as 
\begin{eqnarray}
m_\eta^2\equiv m^{2} (\eta^{\pm}) &=& m_2^{2} + \frac12 \lambda_3 v^{2}
 + \frac12 \lambda_8 v'^{2}, \\ 
m^2_{\eta_{R}}\equiv m^{2} (\mathrm{Re}\:\eta^{0}) &=& m_2^{2} + \frac12 \lambda_8 v'^{2}
 + \frac12 (\lambda_3 + \lambda_4 + 2\lambda_5) v^{2}, \\ 
m^2_{\eta_{I}}\equiv m^{2} (\mathrm{Im}\:\eta^{0}) &=& m_2^{2} + \frac12 \lambda_8 v'^{2}
 + \frac12 (\lambda_3 + \lambda_4 - 2\lambda_5) v^{2},\\
m^2_{\chi_{R}} &=& m_4^{2} + \frac12 \lambda_{10} v^2 + \frac12\lambda_{12} v'^{2} + \sqrt2 m_5v', \\ 
m^2_{\chi_{I}} &=&  m_4^{2} + \frac12 \lambda_{10} v^2 + \frac12\lambda_{12} v'^{2} -\sqrt2 m_5v'.
\end{eqnarray}
The tadpole conditions for $\eta$ and $\chi$, which are given by
$\left.\frac{\partial V}{\partial \eta}\right|_{\mathrm{VEV}}=0$, 
$\left.\frac{\partial V}{\partial \chi}\right|_{\mathrm{VEV}}=0$, 
$0<\left.\frac{\partial^2 V}{\partial \eta^2}\right|_{\mathrm{VEV}}$ and 
$0<\left.\frac{\partial^2 V}{\partial \chi^2}\right|_{\mathrm{VEV}}$ 
tell us that  
\be
0<m^2_2+\frac{v^2}2( \lambda_3+\lambda_4+2\lambda_5)+ \frac{v'^2}2\lambda_8,\quad
0<m^2_4+\frac{v^2}2 \lambda_{10}+ \sqrt2m_5v'+\frac{v'^2}2\lambda_{12},
\ee
in order to satisfy the condition $v_\eta=0$ and $v_\chi=0$ at tree level, respectively.
In order to avoid that $\langle \Phi \rangle=\langle \Sigma \rangle=0$ be 
a local minimum, we require the following condition: 
\be
\lambda_7-\frac{2}{3}\sqrt{\lambda_1\lambda_6}<0.
\ee
To achieve the global minimum  at $\langle \eta \rangle=\langle \chi \rangle=0$, we find the following condition
\be
0<\lambda_{11}-\frac{2}{3}\sqrt{\lambda_2\lambda_9}.
\ee
Finally, if the following conditions  
\bea
&&0<\lambda_{3}+\frac{2}{3}\sqrt{\lambda_1\lambda_2},
\quad 0<\lambda_{7}+\frac{2}{3}\sqrt{\lambda_1\lambda_6},\quad
0<\lambda_{10}+\frac{2}{3}\sqrt{\lambda_1\lambda_9},
\nn\\&&
 0<\lambda_{8}+\frac{2}{3}\sqrt{\lambda_2\lambda_6},\quad
0<\lambda_{11}+\frac{2}{3}\sqrt{\lambda_2\lambda_9},\quad 
0<\lambda_{12}+\frac{2}{3}\sqrt{\lambda_6\lambda_9}, 
\eea
are satisfied, the Higgs potential Eq.(\ref{HP}) is bounded from below. 

\subsection{S and T parameters}
It is worth mentioning the new contributions to the S and T parameters due to the new scalar boson $\eta$,
which are given in Refs.~\cite{Barbieri:2006dq, Peskin:1991sw} as
\bea
{\rm S}_{\mathrm{new}}&=&
\frac{1}{2\pi}\int_0^1 dx (1-x)x\ln \left[\frac{x m_{\eta_R}^2+(1-x)m_{\eta_I}^2}{m^2_{\eta}}\right],
\\
{\rm T}_{\mathrm{new}}&=&
\frac{1}{32\pi^2 \alpha_{{\rm em}} v^2} \Biggl[ F(m_{\eta}, m_{\eta_R})+  F(m_{\eta}, m_{\eta_I})- F(m_{\eta_I}, m_{\eta_R}) \Biggr],\\
F(m_1, m_2)&=&
\frac{m_1^2+m_2^2}{2}-\frac{m_1^2m_2^2}{m_1^2-m_2^2}\ln\left(\frac{m_1^2}{m_2^2}\right),
\eea
where $\alpha_{\mathrm{em}}=1/137$ is the fine structure constant. 
The experimental deviations from the SM predictions,  under
$m_{h_{SM}}=126$ GeV, are given by \cite{Baak:2012kk}
\be
{\rm S}_{\mathrm{new}}=0.03\pm 0.10,\quad {\rm T}_{\mathrm{new}}=0.05\pm 0.12,
\label{STallowed}
\ee
When the masses are $1\leq m_1/m_2\lesssim3$, the function $F(m_1,m_2)$
is approximated to 
\begin{equation}
F\left(m_1,m_2\right)\approx
 \frac{2}{3}\left(m_1-m_2\right)^2,
\label{eq:s-t}
\end{equation}
as in Ref.~\cite{Barbieri:2006dq}. From Eq~(\ref{STallowed}),
we get the following constraint for $\eta$ masses,
\begin{equation}
\left(m_\eta-m_{\eta_R}\right)
\left(m_\eta-m_{\eta_I}\right)\lesssim133~\mathrm{GeV}.
\label{st-cond}
\end{equation}

\subsection{Neutrino mass matrix and LFV processes}
\begin{figure}[t]
\begin{center}
 \includegraphics[scale=1]{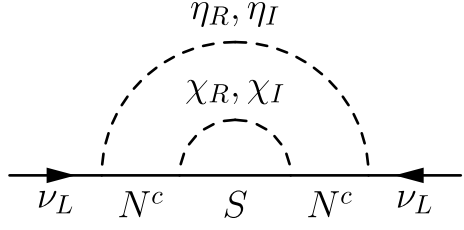}
   \caption{Neutrino mass generation via two-loop radiative seesaw.}
   \label{two-neut}
\end{center}
\end{figure}

The active neutrino mass matrix at the two-loop level as depicted in
Fig.\ref{two-neut} is given by 
\begin{equation}
\left(m_\nu\right)_{\alpha\beta}=
\left(y_{\nu} y_N^*y_N^\dag y_{\nu}^T\right)_{\alpha\beta}
\frac{m_{S}}{4(4\pi)^4}
\int_0^1dx\int_0^{1-x}dy\frac{1}{x(1-x)}
\Biggl[
I\left(m_S^2,m_{RR}^2,m_{RI}^2\right)
-I\left(m_S^2,m_{IR}^2,m_{II}^2\right)
\Biggr],
\label{neu-mass}
\end{equation}
where
\begin{eqnarray}
I(m_1^2,m_2^2,m_3^2)\!\!\!&=&\!\!\!
\frac{m_1^2 m_2^2\log\left(\displaystyle\frac{m_2^2}{m_1^2}\right)
+m_2^2 m_3^2\log\left(\displaystyle\frac{m_3^2}{m_2^2}\right)+
m_3^2 m_1^2\log\left(\displaystyle\frac{m_1^2}{m_3^2}\right)}
{(m_1^2-m_2^2)(m_1^2-m_3^2)}
,\\
m_{ab}^2\!\!\!&=&\!\!\!\frac{ym_{\eta_{a}}^2+xm_{\chi_{b}}^2}{x(1-x)}\quad
(a,b=R\:\mathrm{or}\:I),
\end{eqnarray}
and $m_S$ is the mass of $S$, abbreviating generation index of $S$. 
Since the neutrino mass scale should be roughly 
$m_{\nu}\sim10^{-1}~\mathrm{eV}$, 
the product of $y_N^2y_{\nu}^2$ 
and the integral by $x$ and $y$ of order $10^{-8}$ is required 
when $m_S\sim 1~\mathrm{TeV}$. If $y_N^2y_{\nu}^2 \sim 1$, 
$m_{\eta_R}\simeq m_{\eta_I}$ is required, which is realized by small 
$\lambda_5$.


The Branching Ratio (Br) of charged Lepton Flavor Violation (LFV) processes 
$\ell_{\alpha} \to \ell_{\beta}\gamma~(\alpha,\beta=e,\mu,\tau)$ 
is given by 
\begin{equation}
\mathrm{Br}\left(\ell_\alpha\to\ell_\beta\gamma\right)=
\frac{\alpha_{\mathrm{em}}\left|\left(y_{\nu}y_\nu^\dag\right)_{\alpha\beta}\right|^2}
{768\pi G_F^2m_\eta^4}
\mathrm{Br}\left(\ell_\alpha\to\ell_\beta\nu_\alpha\overline{\nu_\beta}\right),
\label{eq:lfv}
\end{equation}
where right-handed neutrino mass is neglected here. 
The latest limit for $\mu\to e\gamma$ is given by MEG
experiment~\cite{meg2} as
\be
\mathrm{Br}(\mu \to e \gamma)<5.7\times 10^{-13}.
\label{constraints}
\ee
For sum of active neutrino masses, the limit of $\sum
m_{\nu}<0.933~\mathrm{eV}$ is imposed from the cosmological
observation~\cite{Ade:2013lta}. 
In the next section, we take into account these constraints
of S-T parameters,  LFV and the neutrino mass in the discussion of DM. 

\section{Dark Matters}
We discuss the DM properties in this section. 
The $\mathbb{Z}_2$ parity imposed to the model stabilizes DM. In addition
to the $\mathbb{Z}_2$ parity, we have a remnant $\mathbb{Z}_6$ symmetry
after $B-L$ symmetry breaking which stabilizes particles charged under
the $\mathbb{Z}_6$ symmetry as well.\footnote{One straightforwardly
finds the remnant $\mathbb{Z}_6$ symmetry is derived and the charges
are obtained by multiplying $6$ to the $B-L$ charges of all the
particles so as to being the minimal integers. }
As a result, our model has two or three DM candidates simultaneously, 
and the number of DMs depends on mass hierarchy of DM candidates included in
the model. In general we have five DM candidates
which are $N^c$, $S$, $\bar{S}$ as fermionic DMs and $\chi_{R(I)}$,
$\eta_{R(I)}$ as bosonic DMs (For bosonic part, the DM property of
the imaginary part is almost same as that of the real one). Of these, 
three particles can be DMs when decay of a charged particle under the $\mathbb{Z}_2$ and/or $\mathbb{Z}_6$ symmetry
is kinematically forbidden, otherwise we have two DMs. 
The mass of $N^c$ is expected to be somewhat light ($\simeq{\cal O}(10)$ GeV)
because its mass is generated at the one-loop level.
The other particles have typically the mass of $B-L$
symmetry breaking scale. 
Therefore it is natural to choose $N^c$ as the lightest DM. In this
case, $\eta_R$ and $\eta_I$ cannot be DM since they have the same charge
with $N^c$ under $\mathbb{Z}_2$ and $\mathbb{Z}_6$ symmetry. The
remaining DM candidates are $S$, $\bar{S}$, $\chi_{R}$ and $\chi_{I}$. 
The interactions of $S$ and $\bar{S}$ are almost same
except for Yukawa interaction $y_N N^c\chi S$. This Yukawa interaction leads DM-exchanging scattering like $SS\to N^cN^c$. 
The DM candidates $\chi_{R}$ and $\chi_{I}$ have also similar DM-exchanging 
scattering $\chi_{R(I)}\chi_{R(I)}\to N^cN^c$ via Yukawa interaction
$y_N N^c\chi S$ as same as the case of $S$.
Thus as the simplest case, we consider two-component DMs ($N^c$ and
$\bar{S}$) in the following since there is no exchange scattering
between $N^c$ and $\bar{S}$ at the tree level.\footnote{Regarding the
pair of $(N^c,S)$ or $(N^c,\chi_{R(I)})$ as
DMs, it would be the simplest if Yukawa coupling $y_S$ is large enough 
because most of $S$
or $\chi_{R(I)}$ annihilates into $N^cN^c$. In this case, one can
consider that only $N^c$ is substantial DM.} 

\begin{figure}[t]
\begin{center}
\includegraphics[scale=0.8]{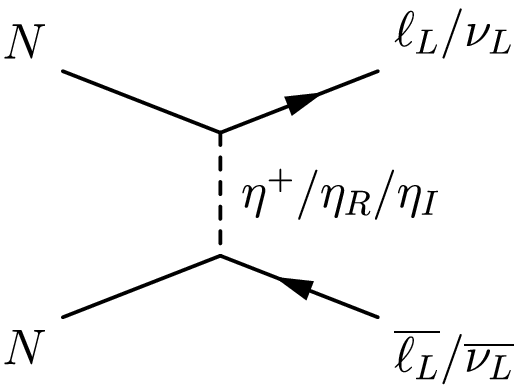}
\quad
\includegraphics[scale=0.8]{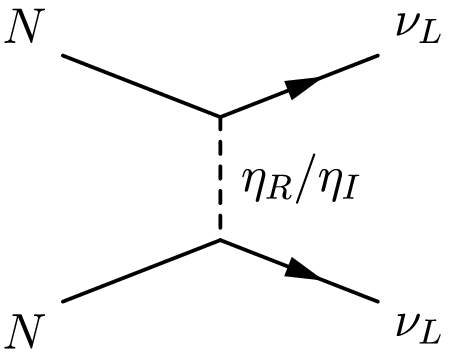}
\quad
\includegraphics[scale=0.8]{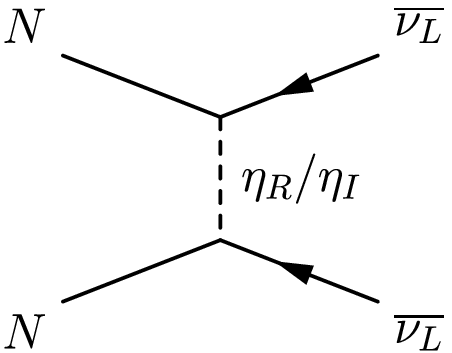}
\\\vspace{0.5cm}
\includegraphics[scale=0.8]{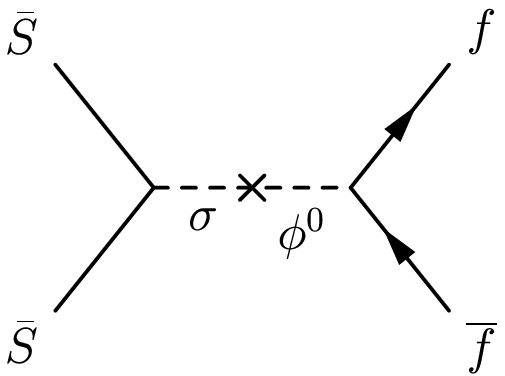}
\quad
\includegraphics[scale=0.8]{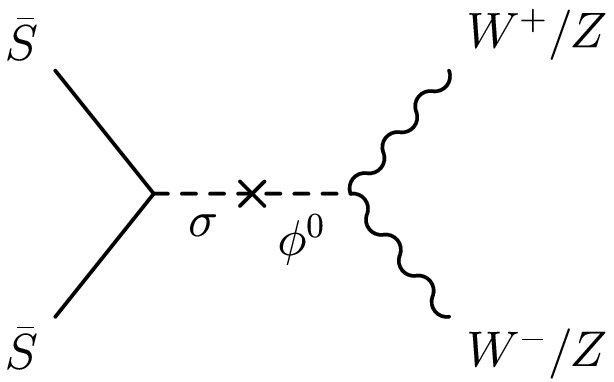}
\quad
\includegraphics[scale=0.8]{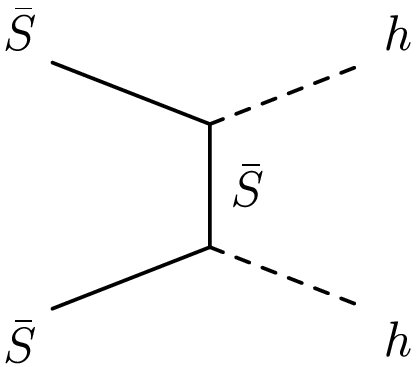}
\caption{The annihilation channels of fermionic DM $N$ (upper row) and
 $\bar{S}$ (lower row).}
   \label{fig:ann}
\end{center}
\end{figure}
The mass matrix of three right-handed neutrinos $N^c$ is radiatively
induced and the expression is found as
\begin{equation}
\left(m_{N^c}\right)_{ij}=\sum_{k=1}^3\frac{(y_N)_{ik}(y_N)_{jk}m_{S_k}}{(4\pi)^2}
\left[\frac{m_{\chi_{R}}^2}{m_{\chi_{R}}^2-m_{{S_k}}^2}\ln\left(\frac{m_{\chi_{R}}^2}{m_{{S_k}}^2}\right)
-\frac{m_{\chi_{I}}^2}{m_{\chi_{I}}^2-m_{{S_k}}^2}\ln\left(\frac{m_{\chi_{I}}^2}{m_{{S_k}}^2}\right)\right].
\label{eq:numass}
\end{equation}
We need multi-pair of $S$ and
$\bar{S}$ to generate three non-zero right-handed neutrino masses,
otherwise only one non-zero mass is generated. 
An interesting feature of the model is the connection between the light
neutrino masses~Eq.~(\ref{neu-mass}) and the right-handed neutrino
masses~Eq.~(\ref{eq:numass}). The right-handed neutrino masses tend to be
small corresponding to the tiny light neutrino masses. 
The mass matrix can be diagonalized by a unitary matrix and the
lightest particle, which is called as simply $N$ below, is DM. We 
choose the diagonal basis of right-handed neutrinos. 
The annihilation channels of the right-handed neutrino DM into the SM
particles are shown in the upper part of Fig.~\ref{fig:ann}, where the
annihilation channel via $B-L$ gauge boson is omitted since the
contribution is small enough due to exchange of heavy $B-L$ gauge boson. 
We obtain the annihilation cross sections as 
\begin{eqnarray}
\sigma{v}_{\mathrm{rel}}\left(NN\to\ell_L\overline{\ell_L}\right)
\!\!\!&=&\!\!\!
\frac{\left[\left(y_{\nu}^\dag y_{\nu}\right)_{11}\right]^2}{48\pi
m_{N}^2}
G\left(\alpha_\eta,\alpha_\eta\right)v_{\mathrm{rel}}^2,
\label{eq:ann-nc1}\\
\sigma{v}_{\mathrm{rel}}\left(NN\to\nu_L\overline{\nu_L}\right)
\!\!\!&=&\!\!\!
\frac{\left[\left(y_{\nu}^\dag y_{\nu}\right)_{11}\right]^2}{48m_N^2}
\frac{1}{4}\sum_{a,b=R,I}G\left(\alpha_a,\alpha_b\right)v_{\mathrm{rel}}^2,
\label{eq:ann-nc2}\\
\sigma{v}_{\mathrm{rel}}\left(NN\to\nu_L\nu_L\right)
\!\!\!&=&\!\!\!
\frac{\left[\left(y_{\nu}^\dag y_{\nu}\right)_{11}\right]^2}{64\pi
m_{N}^2}\left(\alpha_R-\alpha_I\right)^2
\left[
1+\frac{v_{\mathrm{rel}}^2}{6}
\Bigl(
3-10\left(\alpha_R+\alpha_I\right)+6\left(\alpha_R+\alpha_I\right)^2-4\alpha_R\alpha_I
\Bigr)
\right]\nonumber\\
\!\!\!&=&\!\!\!
\sigma{v}_{\mathrm{rel}}\left(NN\to\overline{\nu_L}\overline{\nu_L}\right).
\label{eq:ann-nc3}
\end{eqnarray}
where $G\left(x,y\right)=xy\left(1-x-y+2xy\right)$,
$\alpha_\eta=m_{N}^2/\left(m_{N}^2+m_{\eta}^2\right)$ and
$\alpha_a=m_{N}^2/\left(m_{N}^2+m_{\eta{a}}^2\right)$, $a=R,I$. 
In the last expression, the symmetric factor $1/2$ should be multiplied when 
the flavor of the final state $\nu_L\nu_L$ is the same. 
The chiral suppression is not effective for the pair of (anti-)neutrino final state
channels if the mass splitting between $\eta_R$ and $\eta_I$ is not
negligible. 

We analyze the relic abundance of the lightest right-handed neutrino $N$
with the constraints from S-T parameters, the neutrino mass and LFV
Eq.~(\ref{constraints}). 
We sweep the parameters in the following range:
\begin{eqnarray}
10^2~\mathrm{GeV}<m_S<10^4~\mathrm{GeV},&&
10^{-4}<\left(y_\nu\right)_{e1}\approx\left(y_{\nu}\right)_{\mu1}<1,\\
10^2~\mathrm{GeV}<m_{\eta_{R(I)}}<10^{3}~\mathrm{GeV},
&&
10^2~\mathrm{GeV}<m_{\chi_{R(I)}}<10^{4}~\mathrm{GeV}.
\end{eqnarray}
The mass of $\eta$ must satisfy the constraint of S-T parameters
Eq.~(\ref{st-cond}), and the typical mass scale of $S$, $\chi$ and $\eta$ is of $\cal{O}$(1) $\mathrm{TeV}$ 
since we assume the $B-L$ breaking scale is several $\mathrm{TeV}$. Only the Yukawa couplings $\left(y_{\nu}\right)_{\tau1}$ and
$y_{N}$ are fixed to $\left(y_{\nu}\right)_{\tau1}= 1.0$ and 
$(y_N)_{ik} ={\cal O}(1)$
in order to have a proper annihilation cross section, being consistent with
$\mu\to e\gamma$. 
As mentioned before, even if some elements of $y_{\nu}$ and $y_N$ are 
of ${\cal O}(1)$, one can obtain correct neutrino masses and mixings 
by taking small value of $\lambda_5$ and 
by choosing appropriate texture of Yukawa couplings.
As concerning above factors, the right-handed neutrino masses shown in
Eq.~(\ref{eq:numass})
induced by the one-loop effect become $\mathcal{O}(10)~\mathrm{GeV}$,
otherwise the right-handed neutrino masses become much lighter and the
annihilation cross section will be too small because it is proportional to $\alpha_{\eta,R,I}$. 

\begin{figure}[t]
\begin{center}
\includegraphics[scale=0.7]{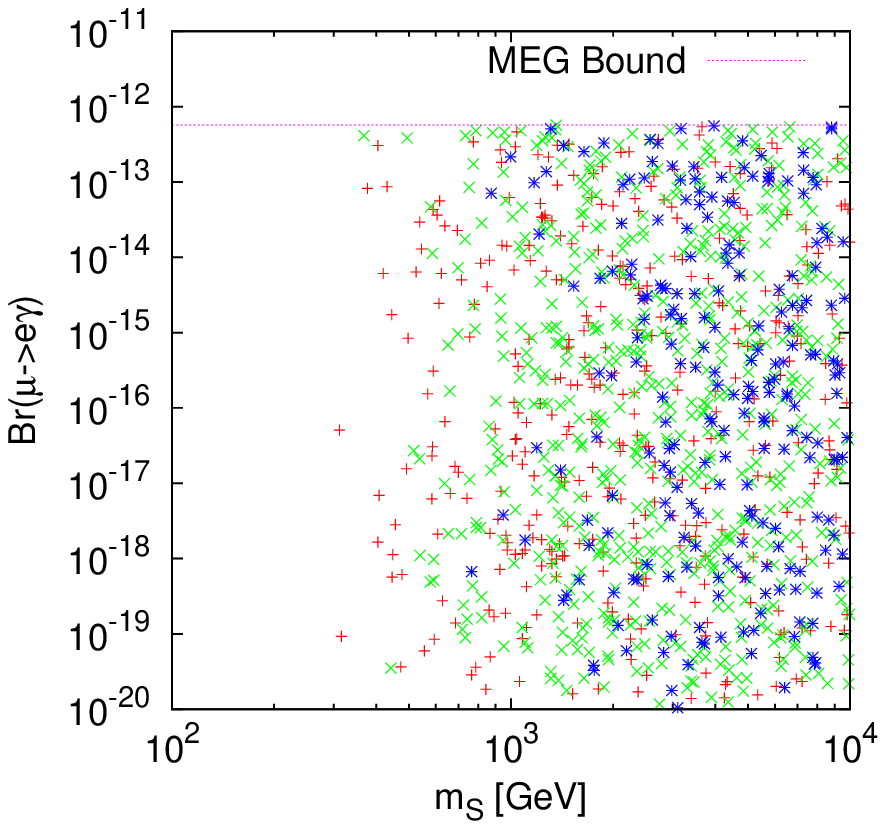}
\includegraphics[scale=0.7]{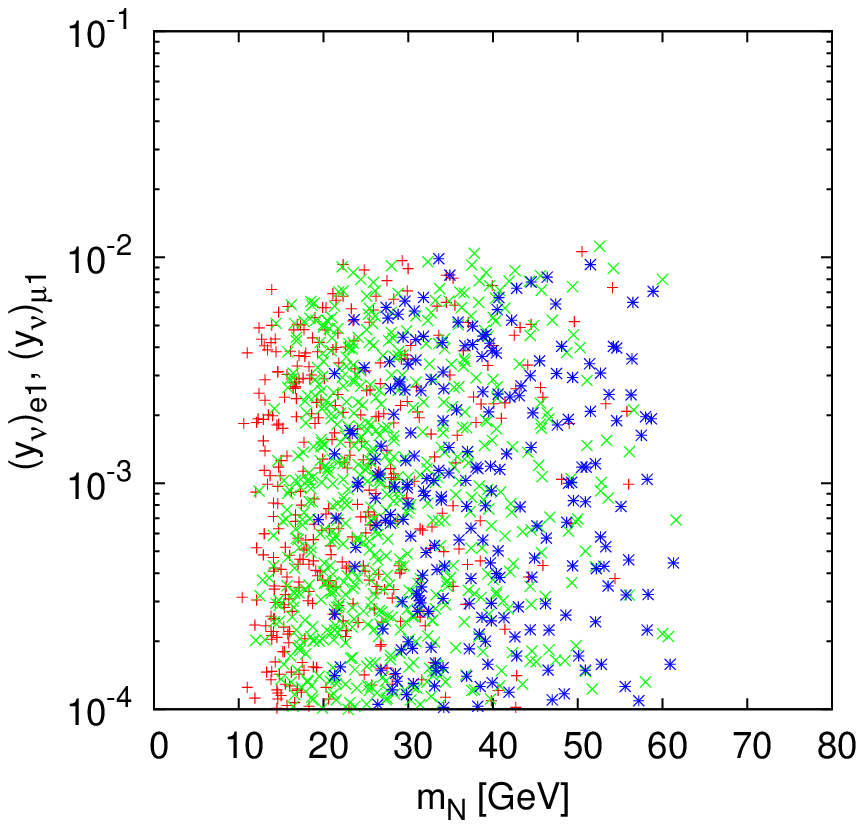}
\includegraphics[scale=0.7]{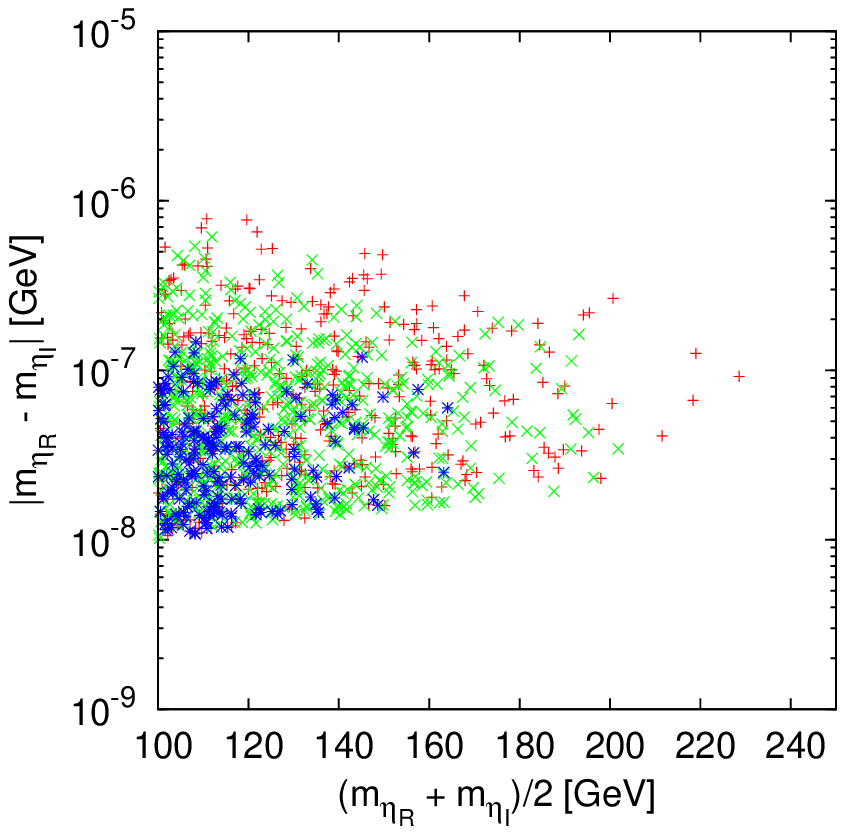}
\includegraphics[scale=0.7]{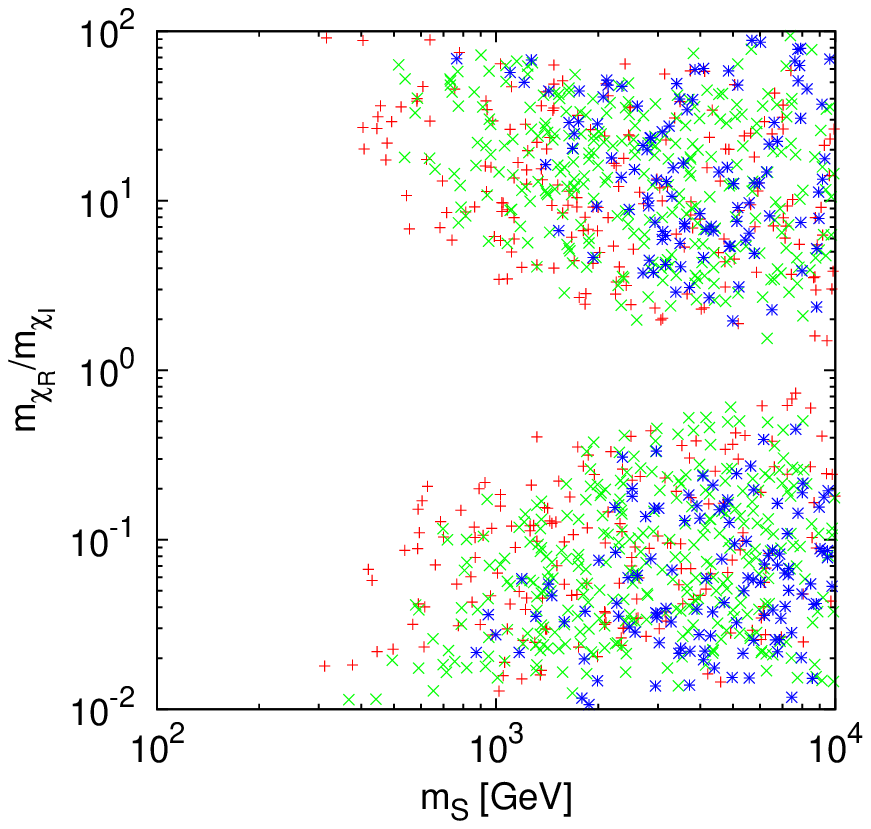}
\caption{The parameter spaces of satisfying LFV, the light neutrino mass
 scale, the thermal DM relic density. The red, green, blue points imply
 the DM relic density fraction of $N$ as $0.7<\Omega_N/\Omega<1.0$,
 $0.3<\Omega_N/\Omega<0.7$, $\Omega_N/\Omega<0.3$ respectively.}
\label{fig:NDM}
\end{center}
\end{figure}

The result in the case of $N$ DM is shown in Fig.~\ref{fig:NDM}, and each point satisfies
the constraint from S-T parameters, appropriate light neutrino mass scale
$\sim10^{-1}~\mathrm{eV}$, the LFV constraint $\mathrm{Br}\left(\mu\to
e\gamma\right)<5.7\times10^{-13}$ and thermal DM relic density
which corresponds to
$\sigma{v}_{\mathrm{rel}}\gtrsim3\times10^{-26}~\mathrm{cm^3/s}$. 
The reason of the inequality in the cross section is that 
we have two DM candidates.  
One should note that total relic density is supplied by $N$ and the other DM
$\bar{S}$. Therefore even if the annihilation cross section is larger than
$3\times10^{-26}~\mathrm{cm^3/s}$, the lack of the amount of DM is
compensated by the amount of $\bar{S}$. 
One can see from the upper left panel that more than around
$300~\mathrm{GeV}$ of $m_S$ can
satisfy the current upper bound of the branching ratio for $\mu\to
e\gamma$. At the same time, Yukawa couplings relating with $\mu\to
e\gamma$ should be less than $10^{-2}$ as shown in the upper
right panel. The right-handed neutrino mass
are of $10\sim80$ $\mathrm{GeV}$ in this case. 
The magnitude of Yukawa couplings is also important to obtain the proper
light neutrino mass scale (See Eq.~(\ref{neu-mass})). In addition to the 
Yukawa coupling, a small mass splitting between $\eta_R$ and $\eta_I$ is 
required to obtain correct light neutrino masses as the left lower panel
shows. As a result, our scenario of DMs is still p-wave dominant since
Eq.~(\ref{eq:ann-nc3}) is suppressed. 
Due to the small $\eta$ mass splitting, we can get tiny neutrino masses 
because of a cancellation among
the integrands of Eq.~(\ref{neu-mass}). 
The masses of $\eta_R$ and $\eta_I$ are 
restricted to be roughly $10^2~\mathrm{GeV}$. Thus charged $\eta^+$ mass
is less than around $300~\mathrm{GeV}$ from the constraint of S-T parameters. 
In addition, charged Higgs search at LHC would give a stringent
bound for $\eta^+$. Some mass regions of charged Higgs for SUSY models
are excluded by the decay of the charged Higgs into slepton plus missing
energy~\cite{ATLAS:2012zka, Aad:2012pxa, CMS:aro}. According to the
CMS result~\cite{CMS:aro}, $120\lesssim
m_{\tilde{\ell}}\lesssim270~\mathrm{GeV}$ is
excluded for slepton mass. 
This model shows a similar signal via $\eta^+\to \ell_\alpha N$, thus
almost same mass region of $\eta^+$ can be expected to be excluded. 
The lower bound of charged Higgs is also obtained from LEP experiment,
and the bound is around 70 GeV under some
conditions~\cite{Swiezewska:2012eh,Lundstrom:2008ai}.
Regarding masses of $\chi_R$ and $\chi_I$ as 
shown in the right-lower panel, mass hierarchy between $\chi_R$ and
$\chi_I$ is necessary to get the scale of
the right-handed neutrino masses which is connected with the size of the
annihilation cross section. 

Next, we move on discussion of $\bar{S}$ DM. This DM does not have any
interactions with $N$ at tree level. Hence we can consider these two DMs 
separately. 
The annihilation cross sections of $\bar{S}$ shown in the lower part of
Fig.~\ref{fig:ann} are given as
\begin{eqnarray}
\sigma{v}_{\mathrm{rel}}\left(\bar{S}\bar{S}\to f\overline{f}\right)
\!\!\!&=&\!\!\!
\sum_{f}\frac{c_fy_{\bar{S}}^2y_f^2}{8\pi}m_{\bar{S}}^2
\left|\frac{1}{D}\right|^2
\left(1-\frac{m_f^2}{m_{\bar{S}}^2}\right)^{3/2}v_{\mathrm{rel}}^2,
\label{eq:ann-s1}\\
\sigma{v}_{\mathrm{rel}}\left(\bar{S}\bar{S}\to W^+W^-\right)
\!\!\!&=&\!\!\!
\frac{y_{\bar{S}}^2g_2^2}{64\pi}m_W^2
\left|\frac{1}{D}\right|^2
\left(3-4\frac{m_{\bar{S}}^2}{m_W^2}+4\frac{m_{\bar{S}}^4}{m_W^4}\right)
\left(1-\frac{m_W^2}{m_{\bar{S}}^2}\right)^{1/2}v_{\mathrm{rel}}^2,
\label{eq:ann-s2}\\
\sigma{v}_{\mathrm{rel}}\left(\bar{S}\bar{S}\to ZZ\right)
\!\!\!&=&\!\!\!
\frac{y_{\bar{S}}^2(g^2+g_2^2)}{128\pi}m_Z^2
\left|\frac{1}{D}\right|^2
\left(3-4\frac{m_{\bar{S}}^2}{m_Z^2}+4\frac{m_{\bar{S}}^4}{m_Z^4}\right)
\left(1-\frac{m_Z^2}{m_{\bar{S}}^2}\right)^{1/2}v_{\mathrm{rel}}^2,
\label{eq:ann-s3}\\
\sigma{v}_{\mathrm{rel}}\left(\bar{S}\bar{S}\to hh\right)
\!\!\!&=&\!\!\!
\frac{y_{\bar{S}}^4\sin^4\alpha}{8\pi m_{\bar{S}}^2}
\left(1-\frac{m_h^2}{m_{\bar{S}}^2}\right)^{1/2}\beta_h^2
\left[1-\frac{\beta_h}{3}\left(1-\frac{m_h^2}{m_{\bar{S}}^2}\right)
+\frac{\beta_h^2}{12}\left(1-\frac{m_h^2}{m_{\bar{S}}^2}\right)^2\right]
v_\mathrm{rel}^2,\nonumber\\
\label{eq:ann-s4}
\end{eqnarray}
where $m_{\bar{S}}$ is the mass of $\bar{S}$ and the color factor $c_f$
is $1$ for leptons and $3$ for quarks. The parameter  
$\beta_h$ is defined as
$\beta_h=m_{\bar{S}}^2/\left(2m_{\bar{S}}^2-m_h^2\right)$ and $D$
is the propagator of the SM-like Higgs $h$ and an extra Higgs $H$,
\begin{equation}
\frac{1}{D}=
\frac{\sin\alpha\cos\alpha}{4m_{\bar{S}}^2-m_h^2+im_h\Gamma_h}-
\frac{\sin\alpha\cos\alpha}{4m_{\bar{S}}^2-m_H^2+im_H\Gamma_H}.
\end{equation}
The $HH$ final state process can be obtained by replacing $m_h\to m_H$
and $\sin\alpha\to\cos\alpha$ in Eq.~(\ref{eq:ann-s4}).

\begin{figure}[t]
\begin{center}
\includegraphics[scale=0.75]{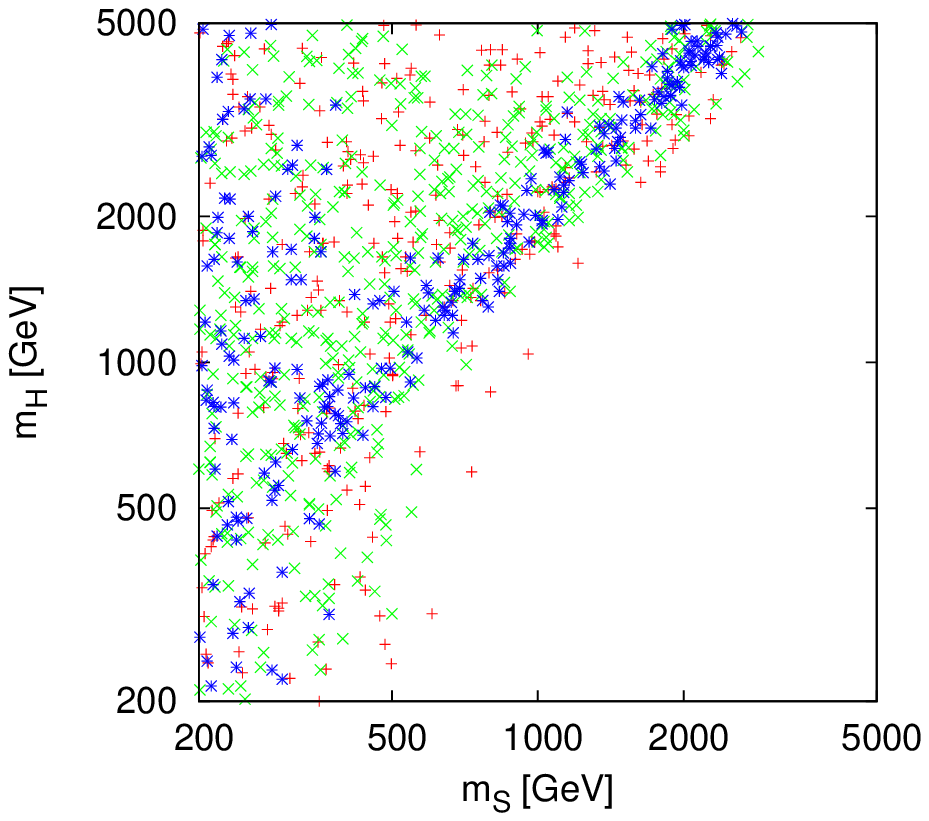}
\includegraphics[scale=0.75]{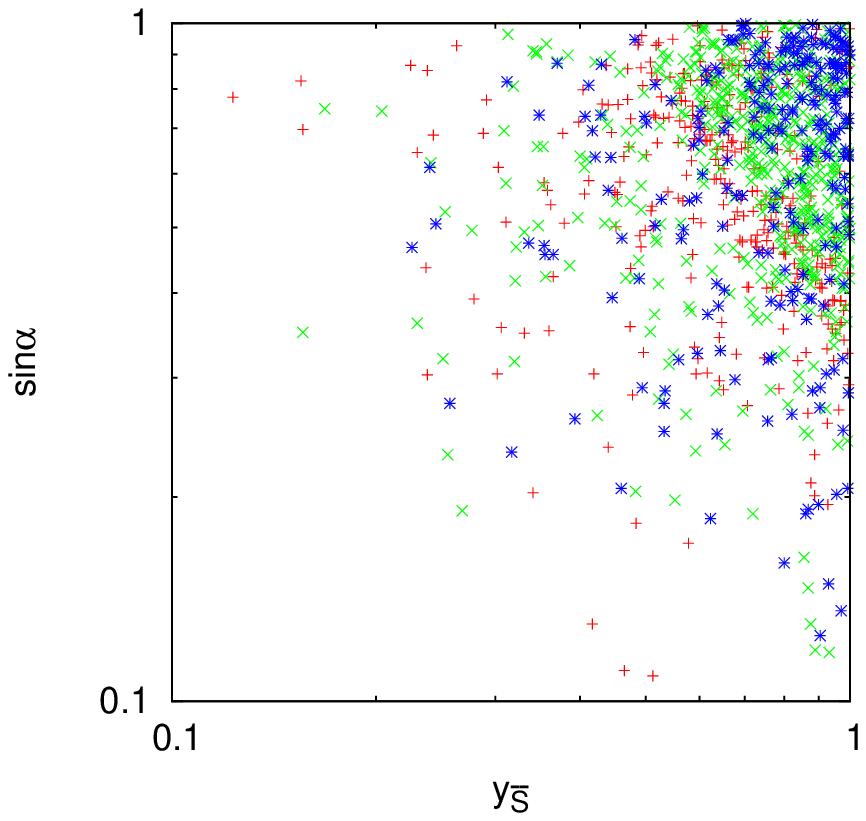}
\caption{The allowed parameter spaces for $\bar{S}$ DM. 
The red, green, blue points imply  the DM relic density fraction of $\bar{S}$
 as $0.7<\Omega_{\bar{S}}/\Omega<1.0$, $0.3<\Omega_{\bar{S}}/\Omega<0.7$,
 $\Omega_{\bar{S}}/\Omega<0.3$ respectively.}
\label{fig:Sbar}
\end{center}
\end{figure}

We have $m_{\bar{S}}$, $y_{\bar{S}}$, $\sin\alpha$, $m_H$ as
parameters and sweep in the following range:  
\begin{eqnarray}
200~\mathrm{GeV}<m_{\bar{S}(H)}<5000~\mathrm{GeV},\quad
10^{-3}<y_{\bar{S}}<1,\quad10^{-3}<\sin\alpha<1,
\end{eqnarray}
satisfying $\sigma{v}\gtrsim3\times10^{-26}~\mathrm{cm^3/s}$.
The result is shown in Fig.~\ref{fig:Sbar}. 
The left panel shows
that the $m_{\bar{S}}<m_H$ roughly holds when $m_{\bar{S}}$ is larger than
around $1~\mathrm{TeV}$. Moreover the couplings have to be of order one to
get correct relic density of $\bar{S}$ (the right panel). It causes the
elastic cross section with nuclei to be larger than the upper bound by
XENON100 as we will discuss below. 

\begin{figure}[t]
\begin{center}
\includegraphics[scale=0.7]{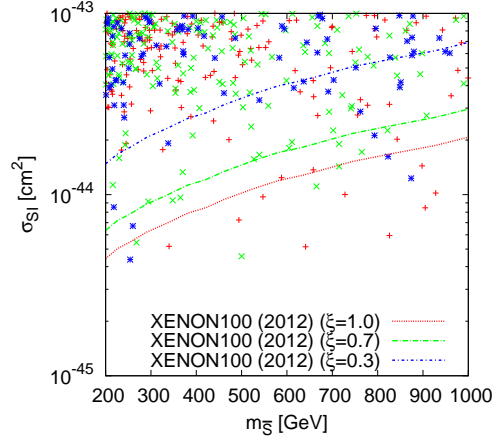}
\caption{The comparison of the elastic cross section of $\bar{S}$ with
 nucleon where the XENON100 upper bound is also drawn together where the
 parameter $\xi$ stands for the fraction of relic abundance of $\bar{S}$ 
 to the total abundance.}
\label{fig:dd}
\end{center}
\end{figure}

Let us move on to the discussion of direct detection of DMs. 
Although our DM consists of two components $N$ and $\bar S$, $N$ does not have any interactions with quarks at tree level
since it is a right-handed neutrino.\footnote{The DM $N$ can interact with quarks at
loop level through electromagnetic interactions~\cite{Schmidt:2012yg, Chang:2010en} and it would be
large as same as being detected by the XENON experiment.}
The other DM $\bar{S}$ interacts with quarks via Higgs exchange. 
Thus it is possible to explore the DM in direct detection experiments
like XENON100 \cite{xenon100}. 
The Spin Independent (SI) elastic cross section $\sigma_{\mathrm{SI}}$
with proton $p$ is given by
\begin{equation}
\sigma_{\mathrm{SI}}=
\frac{4\mu_{\bar{S}}^2}{\pi}\frac{y_{\bar{S}}^2\sin^2\!\alpha\cos^2\!\alpha\:
 m_p^2}{2v^2}
\left(\frac{1}{m_h^2}-\frac{1}{m_H^2}\right)^2
\left(\sum_{q}f_q^p\right)^2,
\label{eq:dd}
\end{equation}
where $\mu_{\bar{S}}=\left(m_{\bar{S}}^{-1}+m_p^{-1}\right)^{-1}$ is the
DM-proton reduced mass. 
The parameters $f_q^p$ which imply the contribution of each quark to
proton mass are calculated by the lattice
simulation~\cite{Corsetti:2000yq, Ohki:2008ff} as
\begin{eqnarray}
&&f_u^p=0.023,\quad
f_d^p=0.032,\quad
f_s^p=0.020,
\end{eqnarray}
for the light quarks and $f_Q^n=2/27\left(1-\sum_{q\leq 3}f_q^n\right)$ for
the heavy quarks $Q$ where $q\leq3$ implies the summation of the light
quarks. The recent another calculation is performed in
Ref.~\cite{Alarcon:2011zs}.
Fig.~\ref{fig:dd} is the comparison with XENON100 upper
bound~\cite{xenon100} with the same parameter obtained from the analysis
of the relic density. In the case that $\bar S$ DM to be dominant,
most of parameter region allowed by the relic density is 
excluded by the XENON100 upper bound due to the large Yukawa coupling
$y_{\bar{S}}$. Despite of such a strong constraint, some allowed
parameter region certainly exists. These parameters imply that
Yukawa coupling $y_{\bar{S}}$ (with large mixing of $\alpha$) is rather 
small and the mass of $m_{\bar{S}}$ is close to a resonance for the
annihilation cross section in Eq.~(\ref{eq:ann-s1})-(\ref{eq:ann-s3}). 
Needless to say, we can easily relax such a situation because of
multi-component DM scenario. Since we have two DMs in the model, the fraction
parameter of relic density $\xi$, which stands for the fraction of relic abundance of $\bar{S}$ 
 to the total abundance, makes the XENON100 limit looser, and
wide allowed parameters appear.

\section{Conclusions}
We have constructed a two-loop radiative  seesaw model with 
local $B-L$ symmetry at the TeV scale that provides neutrino masses. 
We have also studied the multi-component DM properties, 
in which we have two fermionic DMs with different  mass scale; ${\cal O}$(10) GeV for $N$ and  
$100\sim1000$ GeV for $\bar S$. Although $N$ is right-handed 
neutrino, its mass is generated by the one-loop effect. 
We have shown that the allowed mass regions of the particles
$\eta_{R(I)}$, $\chi_{R(I)}$, an extra Higgs $H$, $S$ and Yukawa
couplings constrained by S-T parameters, Br($\mu\to e\ \gamma$)$<$
5.7$\times10^{-13}$, 
$m_\nu\sim$ 0.1 eV, and annihilation cross section of DMs,
in which  for example we found the mass of $\eta_R$ and $\eta_I$ should
be degenerate: $10^{-8}\lesssim|m_{\eta_R}-m_{\eta_I}|\lesssim10^{-6}$
GeV. The upper bound of $m_N$ is around 80 GeV due to the loop-induced mechanism of $m_N$. 
Too light $m_N$ leads too much relic density of $N$ because of 
the small annihilation cross section. 
We have investigated allowed parameter region from direct detection of
$\bar{S}$ DM through the interaction with Higgses. 
Moreover, we have found that the region of the large mixing $\sin\alpha$
will be testable by the exposure of the future XENON experiment. 
Our model would be revealed by the other characteristic evidences 
such as two gamma line signals in cosmic ray coming from the
annihilation of two component DMs.


\section*{Acknowledgments}
H.O. thanks to Prof. Eung-Jin Chun and Dr. Kei Yagyu for fruitful discussion.
Y.K. thanks Korea Institute for Advanced Study for the travel support
and local hospitality during some parts of this work.
T.T. acknowledges support from the European ITN project (FP7-PEOPLE-2011-ITN,
PITN-GA-2011-289442-INVISIBLES). 
The numerical calculations were carried out on SR16000 at  YITP in Kyoto University.

\end{document}